\documentclass[11pt]{article}
\usepackage[utf8]{inputenc}
\usepackage[T1]{fontenc}
\usepackage[a4paper,bindingoffset=0cm,head=15pt,body={17cm,25cm},top=2.5cm]{geometry}
\usepackage{amsmath,amssymb}
\usepackage{graphicx}
\usepackage{color}
\usepackage{cite}
\usepackage{ifthen}
\usepackage{titlesec}
\usepackage{fancyhdr}
\usepackage{booktabs}
\usepackage[english]{babel}
\usepackage{hyperref}
\usepackage{titlecaps}
\usepackage{xparse}

\titleformat{\section}{\normalfont\bfseries}{\thesection}{2ex}{}
\titlespacing*{\section}{0pt}{2.5ex}{1ex}
\titleformat{\subsection}{\normalfont\bfseries}{\thesubsection}{2ex}{}
\titlespacing*{\subsection}{0pt}{2.5ex}{1ex}
\newcommand{\I}{\ensuremath{\mathrm{i}}}
\newcommand{\E}{\ensuremath{\mathrm{e}}}

\newcounter{authorcounter}
\newcounter{institutecounter}

\newcommand{\wavesauthorlist}{}
\newcommand{\wavesaddresslist}{}
\newcommand{\wavesemail}{}
\newcommand{\wavesfootnotes}{}
\newcommand{\wavesauthorpre}{}

\def\theNumberTest#1{%
  \if\relax\detokenize\expandafter{\romannumeral-0#1}\relax
    true%
  \else
    false%
  \fi
}

\NewDocumentCommand{\wavesspeaker}{ O{} O{} m m}{%
    \ifthenelse{\value{authorcounter} > 1}{%
      \renewcommand{\wavesauthorpre}{, }%
    }{%
      \renewcommand{\wavesauthorpre}{}%
    }%
    \ifx\relax#1\relax
      \renewcommand{\wavesfootnotes}{}
    \else   
      \renewcommand{\wavesemail}{$^\ast$Email: #1}%
      \renewcommand{\wavesfootnotes}{, \ast}
    \fi
    \ifthenelse{\equal{\theNumberTest{#4}}{true}}{%
      \edef\wavesauthorlist{\wavesauthorlist%
        \wavesauthorpre{}\underline{#3}$^{#2%
        }$%
      }%
    }{%
      \edef\wavesauthorlist{\wavesauthorlist%
        \wavesauthorpre\underline{#3}$^{\arabic{authorcounter}%
        }$%
      }
      \edef\wavesaddresslist{\wavesaddresslist%
        \par%
        $^{\arabic{authorcounter}}$#4%
      }%
      \stepcounter{authorcounter}%
      \stepcounter{institutecounter}
    }%

    \ifx\relax#2\relax
          \edef\wavesauthorlist{\wavesauthorlist%
        \wavesauthorpre$^{\wavesfootnotes%
        }$%
      }
    \else
    \ifthenelse{\equal{\theNumberTest{#2}}{true}}{%
      \edef\wavesauthorlist{\wavesauthorlist%
        \wavesauthorpre{}$^{,#2\wavesfootnotes%
        }$%
      }%
    }{%
      \edef\wavesauthorlist{\wavesauthorlist%
        \wavesauthorpre$^{,\arabic{institutecounter}\wavesfootnotes%
        }$%
      }
      \edef\wavesaddresslist{\wavesaddresslist%
        \par%
        $^{\arabic{institutecounter}}$#2%
      }%
      \stepcounter{institutecounter}%
    }%
    \fi
  \ignorespaces
}

\NewDocumentCommand{\wavesauthor}{ O{} O{} m m}{%
    \ifthenelse{\value{authorcounter} > 1}{%
      \renewcommand{\wavesauthorpre}{, }%
    }{%
      \renewcommand{\wavesauthorpre}{}%
    }%
    \ifx\relax#1\relax
      \renewcommand{\wavesfootnotes}{}
    \else   
      \renewcommand{\wavesemail}{$^\ast$Email: #1}%
      \renewcommand{\wavesfootnotes}{, \ast}
    \fi%
    \ifthenelse{\equal{\theNumberTest{#4}}{true}}{%
      \edef\wavesauthorlist{\wavesauthorlist%
        \wavesauthorpre{}#3$^{#4%
        }$%
      }%
    }{%
      \edef\wavesauthorlist{\wavesauthorlist%
         \wavesauthorpre{}#3$^{\arabic{institutecounter}%
         }$%
      }
      \edef\wavesaddresslist{\wavesaddresslist%
        \par%
        $^{\arabic{institutecounter}}$#4%
      }%
      \stepcounter{authorcounter}
      \stepcounter{institutecounter}%
    }%

    \ifx\relax#2\relax
          \edef\wavesauthorlist{\wavesauthorlist%
        $^{\wavesfootnotes%
        }$%
      }
    \else
    \ifthenelse{\equal{\theNumberTest{#2}}{true}}{%
      \edef\wavesauthorlist{\wavesauthorlist%
        $^{,#4\wavesfootnotes%
        }$%
      }%
    }{%
      \edef\wavesauthorlist{\wavesauthorlist%
        $^{,\arabic{institutecounter}\wavesfootnotes%
        }$%
      }
      \edef\wavesaddresslist{\wavesaddresslist%
        \par%
        $^{\arabic{institutecounter}}$#2%
      }%
      \stepcounter{institutecounter}%
    }%
    \fi
  \ignorespaces
}

\fancypagestyle{plain}{%
  \fancyhead{}
}

\newenvironment{wavespaper}[3]{%
  \renewcommand{\wavesauthorlist}{}%
  \renewcommand{\wavesemail}{}%
  \setcounter{authorcounter}{1}%
  \setcounter{institutecounter}{1}%
     #2
  \twocolumn[
    \begin{center}
     \bfseries
     #1
     \bigskip

     \wavesauthorlist
     \mdseries
     \smallskip

     \wavesaddresslist
     \smallskip
 
     \wavesemail
    \end{center}%
  ]
}{%
}

\usepackage{adjustbox}
\usepackage{subcaption}
\usepackage{subcaption}

\newcommand{\br}[0]{\mathbf{r}}
\newcommand{\bk}[0]{\mathbf{k}}
\newcommand{\bu}[0]{\mathbf{u}}

\begin{document}

\begin{wavespaper}{%
Observations of eigenfunctions of solar inertial modes using local correlation tracking of magnetic features
}{%
  \wavesspeaker[joshin@mps.mpg.de]{Neelanchal Joshi}{Max-Planck-Institut f\"ur Sonnensystemforschung, 
37077 G\"ottingen}
  \wavesauthor{Zhi-Chao Liang}{1}
  \wavesauthor{Damien Fournier}{1}
  \wavesauthor[][Institut f\"ur Astrophysik, Georg-August-Universit\"at G\"ottingen, 37077 G\"ottingen]{Laurent Gizon}{1}
}

\section*{Abstract}
Solar inertial modes are quasi-toroidal modes of the Sun that are of practical interest as they allow probing the deep convection zone. Since 2010, solar images of the photospheric magnetic field are made available by HMI onboard the Solar Dynamics Observatory. In this work, we track the motion of the small magnetic features  using a cross correlation technique. Under the assumption that these features are passive tracers, we obtain time series of the horizontal flow field on the solar surface. 
A Singular Value Decomposition is then applied to these data to extract the latitudinal profile as well as the time modulation of the modes of oscillation.

\smallskip\noindent\textbf{Keywords:}
inertial modes, near-surface flows

\section{Introduction}


Inertial modes have been recently discovered at the surface of the Sun whose properties are sensitive to the physical conditions deep in the convection zone \cite{gizon2021}.
Previously, local correlation tracking (LCT) of granulation and time-distance helioseismology has been used for detecting Rossby modes in the Sun \cite{loeptien2018,liang2019}, and ring-diagram analysis has been used to identify different classes of inertial modes \cite{gizon2021}.
Here, we extend the LCT technique to other tracers (magnetic features) to obtain the eigenfunction of inertial modes.

\section{Local Correlation Tracking} \label{sec:lct}

We employ a method known as local correlation tracking to measure flows on the surface of the sun. Features such as inter-granular magnetic networks (see Fig.~\ref{mf}) act as tracers for the underlying flow \cite{november1988}. Let us denote $B(\br,t)$ the line-of-sight component of the magnetic field at spatial position $\br$ and time $t$. We then consider $B_{\theta,\phi}(\br,t)$ obtained by multiplying $B$ by an apodization function centered around colatitude $\theta$ and longitude $\phi$ (green circle in Fig.~\ref{mf}).


\begin{figure}[ht]
  \includegraphics[width=\columnwidth]{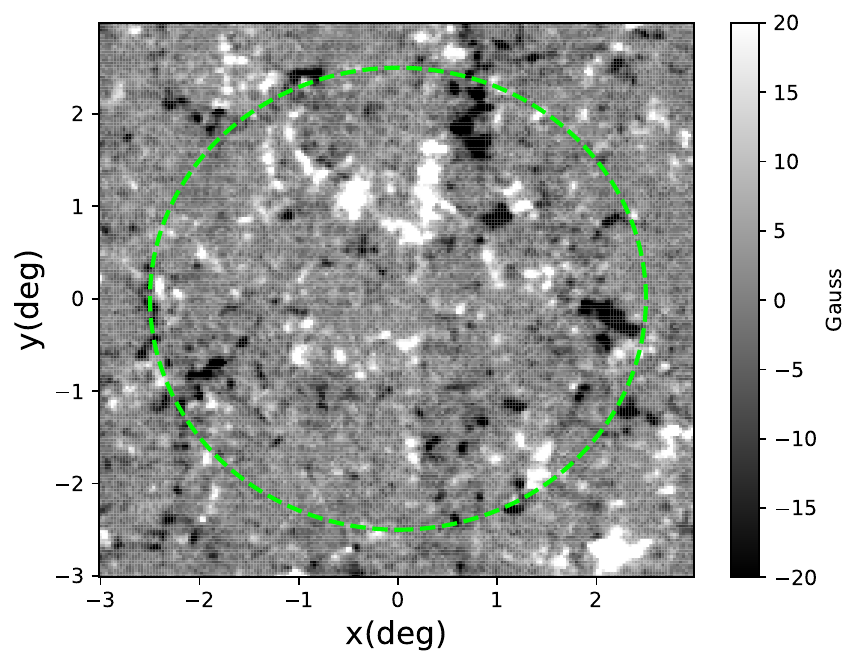}
  \caption{Line-of-sight magnetic field around the equator from HMI magnetogram on 2018.03.25 08:00:00 TAI. The green circle outlines the outer edge of the apodization function.}
  \label{mf}
\end{figure}
For a given target location ($\theta,\phi$), we compute the spatial cross-covariance between two frames  separated by $\delta t$. Equivalently, 
\begin{equation*}
    C_{\theta,\phi}(\delta\br, t) = \int B^*_{\theta,\phi}(\bk,t) B_{\theta,\phi}(\bk, t+\delta t) e^{\mathrm{i} \bk \cdot \delta\br} \mathrm{d}^2\bk , 
\end{equation*}
where $\delta \br$ is the spatial shift between the two images.
Assuming that $\delta t$ is small compared to the evolution time of the features, the velocity $\bu$ is obtained by finding the $\delta\br$ that maximizes the cross-covariance function. Prior to this, the cross-covariance is  averaged over $N$ consecutive realizations  separated by $\Delta t$ to reduce noise: 
\begin{equation*}
    \bu(\theta,\phi,t) = \frac{1}{\delta t} \textrm{argmax}_{\delta\br} \sum_{j=0}^{N-1} C_{\theta,\phi}(\delta\br, t + j\Delta t).
\end{equation*} We fit a parabolic surface around the maximum of the mean cross-covariance function to  find $\delta\br$.


\section{Data description and analysis}
The SDO/HMI instrument has been providing high cadence images of the Sun since May 2010 \cite{schou_hmi}. We use the magnetograms at a cadence of $\delta t =1$~h between May 2010 and Sep 2020 for our analysis. The images are tracked at the Carrington rotation rate to remove the main component of solar rotation with a Postel projection for each apodized region.

We choose $N=6$ and $\Delta t = 1$~h.
We repeat the procedure in \S~\ref{sec:lct} at intervals of $2.5$ heliographic degrees in latitude and longitude to create the flow maps. The resulting $\bu(\theta,\phi,t)$ is at a cadence of 6 hours.


\section{Eigenfunctions of inertial modes}

We obtain $\bu(\theta, \omega, m)$ by Fourier transforming  $\bu(\theta,\phi,t)$ in time and longitude in the Carrington frame.
At fixed $m$, we compute the power spectrum $|\bu(\theta,\omega, m)|^2$. Figure~\ref{ps} shows the power spectrum averaged at high-latitudes ($|\theta - 90^\circ| \ge 60^\circ$) for $m=1$, which displays an excess of power around $-86$~nHz corresponding to a high-latitude mode.  LCT provides an improvement over ring-diagram analysis by having a larger disk coverage and a higher signal-to-noise ratio at higher latitudes. 
\begin{figure}[ht]
  \includegraphics[width=\columnwidth]{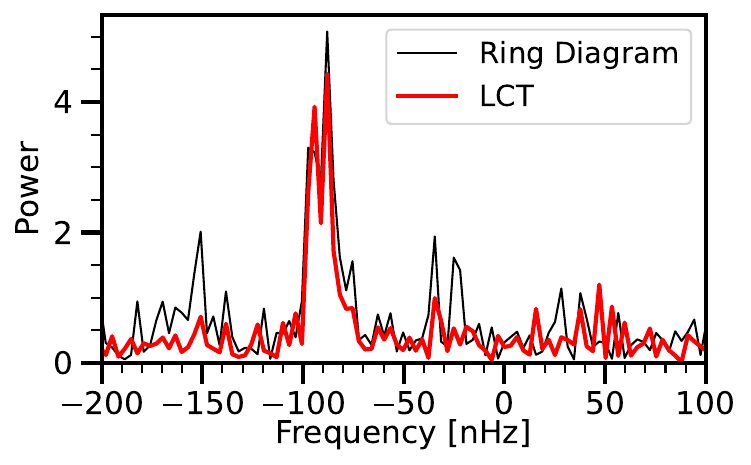}
  \caption{Latitudinally averaged ($|\theta - 90^\circ| \ge 60^\circ$) power spectra of the east-west flow component at $m=1$, from two different methods. The excess power around $-86$~nHz shows the $m=1$ high-latitude mode.}
  \label{ps}
\end{figure}

To obtain the eigenfunction, we filter around the mode frequency and inverse Fourier transform to obtain $\hat{\bu}_{\rm mode}(\theta, t)$.
We get the final latitudinal eigenfunction by performing a Singular Value Decomposition on $\hat{\bu}_{\rm mode}(\theta, t)$ in latitude and time to separate the latitudinal and time dependence of the mode and keep only the first singular value to reduce the noise level $\hat{\bu}_{\rm mode}(\theta, t) \approx S_0 U_0(t)V_0(\theta)$. 
Figure~\ref{uphi_2d} shows the reconstructed 2d eigenfunction $S_0 U_{\rm rms} V_0(\theta) \E^{\I m \phi}$ for the $u_\phi$ component of $m=1$ high-latitude mode, with $U_{\rm rms} = \langle |U_0(t)|^2\rangle^{1/2}_t$. The shape and amplitude is comparable to the one from \cite{gizon2021} using ring-diagram analysis but the spatial coverage extends to higher latitudes.

\begin{figure}[ht]
  \includegraphics[width=\columnwidth]{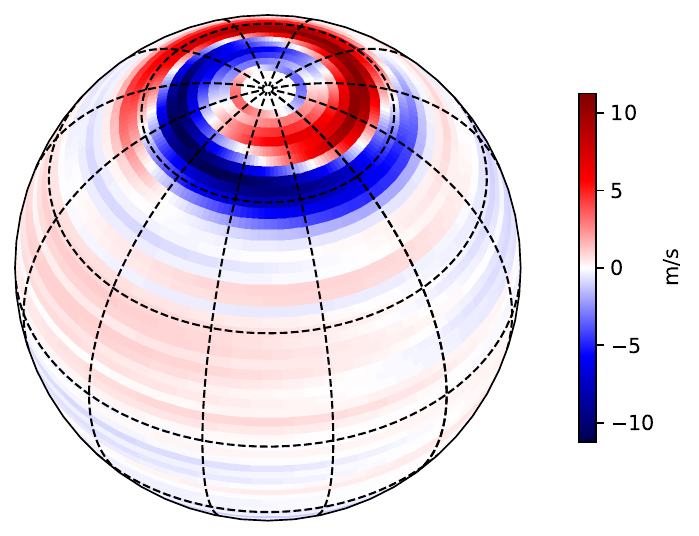}
  \caption{Eigenfunction of $u_\phi$ component of the $m=1$ high-latitude mode obtained from LCT of magnetic network elements}
  \label{uphi_2d}
\end{figure}

This procedure allows us to perform eigenfunction extractions for all classes of inertial modes. 
A detailed comparison of systematic effects and signal-to-noise ratio between the different methods will be the purpose of further studies. 




\end{wavespaper}

\end{document}